# High Order Disturbance Rejection and Tracking via Delayed Feedback Control Method

Zahed Dastan, Mahsan Tavakoli-Kakhki

*Abstract*— **Delayed feedback control is an easy realizable control method which generates control force by comparing the current and the delayed version of the system states. In this paper, a new form of the delayed feedback structure is introduced. Based on the proposed delayed feedback method, a new robust tracking system is designed. This tracking system improves the features of the conventional state feedback with integral action and it is also able to reject higher order disturbances compared to the conventional method. In addition, the proposed tracking system tracks the ramp-shape reference input signal as well, which this is not possible through the conventional state feedback. Due to easy implementable feature of the proposed delayed feedback tracking system, it can be used in practical applications effectively. Moreover, since the proposed method adds delays to the closed loop system dynamics, the ordinary differential equation of the system changes to a delay differential equation with an infinite number of characteristic roots. Thus, conventional pole placement procedures cannot be used to design the delayed feedback controller parameters and place the unstable roots in the left half plane. In this paper, the simulated annealing algorithm is used to determine the proposed control system parameters and move the unstable roots of the delay differential equation to the left half plane. Finally, the efficiency of the proposed reference input tracker is demonstrated on a case study.**

*Index Terms*— **Delayed feedback, disturbance rejection, integral control, robust tracking**

## I. INTRODUCTION

In recent years, the effects of time delays on different systems behavior have been widely taken into consideration. This is due to the existence of delays in many types of systems, including biological [1], economic [2], [3], engineering [4] and social systems [5]. A public view in the field of control system engineering is that delays are always harmful and they may cause instability or decrease the stability margins. It might be asked that whether it is possible to use delays in order to stabilize or improve the system performance or not. An attempt to respond this question is time delay control or TDC. This method at first was proposed to overcome uncertainties in system dynamics [6]. TDC is used in a variety of contexts [7], [8], [9], [10], [11], [12]. Furthermore, time delays can be used as time delay filters [13], [14], [15] and time delay observers [16], [17].

Another special form of utilizing delays in the system control methods is delayed feedback control or time-delay auto synchronization. The main idea of using delayed feedback was proposed by Pyragas in the field of stabilizing unstable oscillations in chaotic systems [18]. In this method, the control force is created by the difference between the current and the delayed value of the system states, e.g., $x(t) - x(t-\tau)$. Delayed feedback has been used in different applications such as engineering [19], chemistry [20], [21] and medicine [22]. This type of feedback has several advantages. Firstly, it can be implemented easily. Secondly, this method is non-invasive. It means that when the desired state is achieved, the control input vanishes. In addition, delayed feedback has been used as an approximation of derivative feedback and phase synchronizer [23], [24]. However, the delayed feedback is superior to the derivative feedback since obtaining reliable measurements of the pure derivative feedbacks in a noisy environment is difficult [25]. Detailed information about the delayed feedback control method can be found in [26].

Delayed feedback can stabilize fixed points and steady states as well as periodic orbits [27]. As a result, its applications are beyond stabilizing periodic orbits. Delays would be exploited to improve the system performance as a tracker. For instance, in [28] the tracking error has been considered as a performance criterion and it has been shown that the existence of delay in the feedback loop decreases the steady state tracking error in the certain types of systems.

Although delays can be used in control systems effectively, by emerging delay terms in the system equations, it is laborious to find the appropriate values of the controller parameters in order to stabilize the unstable system. Several methods have been proposed in the literature to stabilize systems that involve delays in their equations. These methods would be classified to frequency domain tests such as matrix pencil [29] or time domain techniques that utilize Lyapunov's direct method [30]. Detailed discussions about these methods can be found in [31]. For instance, in [32] a delay dependent LMI method has been proposed to examine the stability of a delayed system. Although this method determines a delay interval for which the system can be stabilized, in most cases, designing the controller parameters through similar methods could arise a nonlinear matrix inequality which obviously requires a lot of effort to solve. On the other hand, the frequency domain techniques mostly set a bound on the delay values for which the system is stable. However, these techniques normally do not represent a direct procedure to determine the controller parameters [31]. Consequently, any control method which tries to integrate delays with its structure should tackle these quandaries, and if it is possible, introduces an effective way to find appropriate values for the parameters.

The contribution of the paper is to use delayed feedback characteristics to improve tracking and disturbance rejection

Zahed Dastan is with the Faculty of Electrical Engineering, K. N. University of Technology, Tehran, Iran (email: zaheddastan@ee.kntu.ac.ir)
Mahsan Tavakoli-Kakhki is with the Faculty of Electrical Engineering, K. N. University of Technology, Tehran, Iran (email: matavakoli@kntu.ac.ir)

properties of a tracking system. In this paper, a new delayed feedback structure is proposed to design a robust tracking system based on integral action [33]. It is proved that the proposed control system not only tracks the step/ramp reference input but also an unstable plant can be stabilized by using this control system as well. Most of the proposed methods in the context of the disturbance rejection benefit from observers in order to eliminate the disturbance effects on the control system output [34], [35], [36], [37], [38], [39]. Whereas, due to the particular form of the proposed delayed feedback structure in this paper, the closed loop system is able to reject higher order disturbances on the system state sensors without employing observers. In addition, it is demonstrated that the non-invasive feature of the delayed feedback can be exploited to make the tracking system robust against ramp disturbances on the system dynamics and simultaneously it is possible to track ramp reference inputs precisely. Moreover, because of the aforementioned difficulties to design a controller involving delay terms, utilizing the simulated annealing algorithm, a straightforward method is proposed for determining the delayed feedback controller parameters.

This paper is organized as follows. In section II, the conventional tracking system based on the state feedback with integral action is presented. The proposed tracking system is designed in section III and the simulated annealing algorithm is introduced in order to determine the proposed system parameters. Robustness and tracking characteristics of the proposed system is investigated in section IV and V respectively. Section VI illustrates a numerical example of the proposed system and the paper ends with conclusions drawn in Section VII.

## II. Preliminaries

State feedback with integral action is an ordinary reference tracking approach in the state space model. This tracking approach eliminates the effects of the step-type disturbances on the system output and it is almost robust against the changes of the system parameters. This tracking method has been fully described in [33]. Consider the state space model of a controllable system as follows.

$$\begin{cases} \dot{x}(t) = Ax(t) + Bu(t), \\ y = Cx(t). \end{cases} \tag{1}$$

where $x(t) \in \mathbb{R}^n$ , $A \in \mathbb{R}^{n\times n}$ , $B \in \mathbb{R}^{n\times r}$ , $C \in \mathbb{R}^{m\times n}$ and $y \in \mathbb{R}^m$ . The integral state $q(t) \in \mathbb{R}^m$ is considered such that (2) holds.

$$\dot{q}(t) = y(t) - r(t) = Cx(t) - r(t), \tag{2}$$

where $r(t)$ stands for the reference input. Considering the integral state $q(t)$, the state space model of the augmented system is obtained as

$$\begin{cases} \begin{bmatrix} \dot{x}(t) \\ \dot{q}(t) \end{bmatrix} = \begin{bmatrix} A & 0 \\ C & 0 \end{bmatrix} \begin{bmatrix} x(t) \\ q(t) \end{bmatrix} + \begin{bmatrix} B \\ 0 \end{bmatrix} u(t) + \begin{bmatrix} 0 \\ -I \end{bmatrix} r, \\ y(t) = \begin{bmatrix} C & 0 \end{bmatrix} \begin{bmatrix} x(t) \\ q(t) \end{bmatrix}. \end{cases} \tag{3}$$

In conventional method, it is shown that for a controllable system there is a state feedback $u(t) = -\begin{bmatrix} K' & K'' \end{bmatrix} \begin{bmatrix} x(t) \\ q(t) \end{bmatrix}$ such that the closed loop system becomes stable and for properly chosen values of the coefficients $K'$ and $K''$, the integral sate $q(t)$ goes to a steady state. In other words, the following relation holds.

$$\lim_{t\to\infty} \dot{q}(t) = \lim_{t\to\infty} \left( y(t) - r \right) = 0, \tag{4}$$

which results in

$$\lim_{t\to\infty} y(t) = r. \tag{5}$$

Such a control system rejects the effect of the step-type disturbances on the system output. The block diagram of an integral control system is depicted in Fig. 1.

## III. Reference Tracking by Delayed Feedback Controller

In this section, the tracking system with the delayed feedback control structure is designed. Using the non-invasive feature of the delayed feedback, the integral state $q(t)$ is modified and finally an algorithm is proposed to determine the proposed delayed feedback control structure parameters.

### A. Stabilizing augmented system

In this subsection, instead of using the conventional state feedback for stabilizing the augmented system described in section 2, a new form of the delayed feedback controller is used to stabilize the augmented system.

To this end, the integral state $q(t)$ is considered for which relation (2) holds and it is proposed that the control signal $u(t) \in \mathbb{R}^{r\times 1}$ is applied as follows.

$$u(t) = -\left[ K \left( \sum_{i=0}^{p} \left( \binom{p}{i} (-1)^i x(t - i\tau) \right) \right) + K_1 q(t) \right], \tag{6}$$

where the coefficients $K \in \mathbb{R}^{r\times n}$ , $K_1 \in \mathbb{R}^{r\times m}$ and the delay value $\tau > 0$ are chosen such that the delayed closed loop augmented system becomes stable. Substituting (6) in (3) the state space model of the delayed closed loop augmented system is obtained as (7).

$$\begin{cases}\begin{bmatrix}\dot{x}(t)\\ \dot{q}(t)\end{bmatrix}=\begin{bmatrix}A-BK & -BK_1\\ C & 0\end{bmatrix}\begin{bmatrix}x(t)\\ q(t)\end{bmatrix}+\\ \begin{bmatrix}-BK & 0\\ 0 & 0\end{bmatrix}\left(\sum_{i=1}^{p}\left(\begin{pmatrix}p\\ i\end{pmatrix}(-1)^i\begin{bmatrix}x(t-i\tau)\\ q(t-i\tau)\end{bmatrix}\right)\right)+\begin{bmatrix}0\\ -I\end{bmatrix}r,\\ y(t)=\begin{bmatrix}C & 0\end{bmatrix}\begin{bmatrix}x(t)\\ q(t)\end{bmatrix}.\end{cases} \tag{7}$$

Choosing $u(t)$ as (6), which is a new form of delayed feedback, does not change the non-invasive feature of the delayed feedback structure since as $x(t)$ reaches to the steady state the value of $\sum_{i=0}^{p}\left(\begin{pmatrix}p\\ i\end{pmatrix}(-1)^i x(t-i\tau)\right)$ goes to zero. Moreover, this new form of the delayed feedback structure leads to properly disturbance rejection by the closed loop system. This case will be discussed later.

*B. Modified form of the integral state*

As it was mentioned, delayed feedback is a non-invasive control method. Using this feature, in this paper it is proposed to add an extra term to $\dot{q}(t)$. This extra term brings noticeable interests to the closed loop system which are elaborated in the following sections. The proposed $\dot{q}(t)$ is as follows.

$$\dot{q}(t)=Cx(t)-r-K_2(q(t)-q(t-\tau_q)), \tag{8}$$

Where the coefficient $K_2\in\mathbb{R}^{m\times m}$ can be chosen properly by the designer. Modifying $\dot{q}(t)$ as (8), does not change the steady state value of the system output. Because when the system output reaches to its steady state, $q(t)$ will be equal to $q(t-\tau_q)$ and the extra term in (8) will be eliminated. The advantages of adding this new term to $\dot{q}(t)$ are higher order disturbance rejection and tracking the ramp shape reference inputs by the closed loop system. Eventually, by considering the integral state $q(t)$ in (8), the state space model of the modified closed loop system is obtained as

$$\begin{cases}\begin{bmatrix}\dot{x}(t)\\ \dot{q}(t)\end{bmatrix}=\begin{bmatrix}A-BK & -BK_1\\ C & -K_2\end{bmatrix}\begin{bmatrix}x(t)\\ q(t)\end{bmatrix}\\ +\begin{bmatrix}-BK & 0\\ 0 & 0\end{bmatrix}\left(\sum_{i=1}^{p}\left(\begin{pmatrix}p\\ i\end{pmatrix}(-1)^i\begin{bmatrix}x(t-i\tau)\\ q(t-i\tau)\end{bmatrix}\right)\right)\\ +\begin{bmatrix}0 & 0\\ 0 & K_2\end{bmatrix}\begin{bmatrix}x(t-\tau_q)\\ q(t-\tau_q)\end{bmatrix}+\begin{bmatrix}0\\ -I\end{bmatrix}r,\\ y(t)=\begin{bmatrix}C & 0\end{bmatrix}\begin{bmatrix}x(t)\\ q(t)\end{bmatrix}.\end{cases} \tag{9}$$

Accordingly, the parameters $(K,K_1,K_2,\tau,\tau_q)$ are designed such that (9) is stabled and at the same time the reference tracking objective is fulfilled. The proposed delayed feedback tracker with the modified form of the integral state is depicted in Fig. 2.

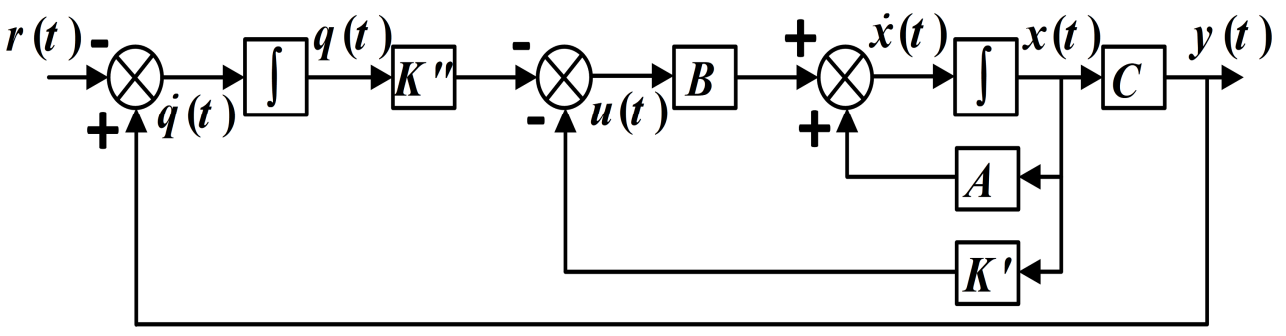

Fig. 1. Block diagram of the conventional state feedback with integral action.

*C. Designing procedure for the controller parameters $(K,K_1,K_2,\tau,\tau_q)$*

In this subsection, it will be discussed that how the parameters of the proposed delayed feedback control structure are determined in order to stabilize the closed loop system and track the reference input. The system equation (9) can be rewritten as follows.

$$\dot{G}(t)=A_0G(t)+A_1\left(\sum_{i=1}^{p}\left(\begin{pmatrix}p\\ i\end{pmatrix}(-1)^i G(t-i\tau)\right)\right)+ A_2G(t-\tau_q)+Tr(t), \tag{10}$$

where

$$G(t)\triangleq\begin{bmatrix}x(t)\\ q(t)\end{bmatrix}, A_0\triangleq\begin{bmatrix}A-BK & -BK_1\\ C & -K_2\end{bmatrix}, A_1\triangleq\begin{bmatrix}-BK & 0\\ 0 & 0\end{bmatrix}$$

$$A_2=\begin{bmatrix}0 & 0\\ 0 & K_2\end{bmatrix}, T=\begin{bmatrix}0\\ -I\end{bmatrix}.$$

The characteristic equation of the system equation (10) is given by

$$\det\left(sI-A_0-A_1\left(\sum_{i=1}^{p}\left(\begin{pmatrix}p\\ i\end{pmatrix}(-1)^i e^{-si\tau}\right)\right)-A_2e^{-s\tau_q}\right)=0. \tag{11}$$

By defining $E(p,i)\triangleq\sum_{i=1}^{p}\left(\begin{pmatrix}p\\ i\end{pmatrix}(-1)^i\right)$, the characteristic equation (11) can be rewritten in the following compact form

$$\det\left(sI-A_0-A_1E(p,i)e^{-si\tau}-A_2e^{-s\tau_q}\right)=0. \tag{12}$$

The primary objective is to determine the parameters of the delayed feedback structure such that all the roots of the characteristic equation (12) lie in the left half plane. As it was mentioned before, usually determining the parameters of the delayed feedback control structure through the available methods for the stability of delayed systems requires a lot of effort. However, in [40] a method has been proposed called continuous pole placement which is based on the sensitivity of $m$ rightmost roots with respect to the controller parameters.

This method moves $m$ rightmost roots of the characteristic equation to the left half plane by applying small perturbations to the controller parameters. The small changes of the controller parameters are obtained from the following equation.

$$\Delta K = S_m^{\dagger} \Delta\Lambda_m^d, \tag{13}$$

$\Delta K = [\Delta k_1 \ \cdots \ \Delta k_n]^T$ indicates the small displacement of the controller parameters, $\lambda_i$ is the characteristic root , $S_m = [s_{i,j}] \in R^{m\times n}, s_{i,j} = \frac{\partial \lambda_i}{\partial k_j}$ is the sensitivity matrix, $S_m^{\dagger}$ is the Moore–Penrose inverse of the $S_m$ and finally $\Delta\Lambda_m^d = [\Delta\lambda_1^d \ \cdots \ \Delta\lambda_m^d]^T$ is small displacement of the $m$ rightmost roots. This algorithm decreases the real part of the rightmost roots until two roots coincide and the sensitivity matrix becomes large. In essence, in [40] there are no evidences to indicate that while this algorithm breaks down, the rightmost roots are in the left half plane. In order to tackle this limitation, it is straightforward to use a numerical algorithm which is able to avoid getting stuck in local minima.

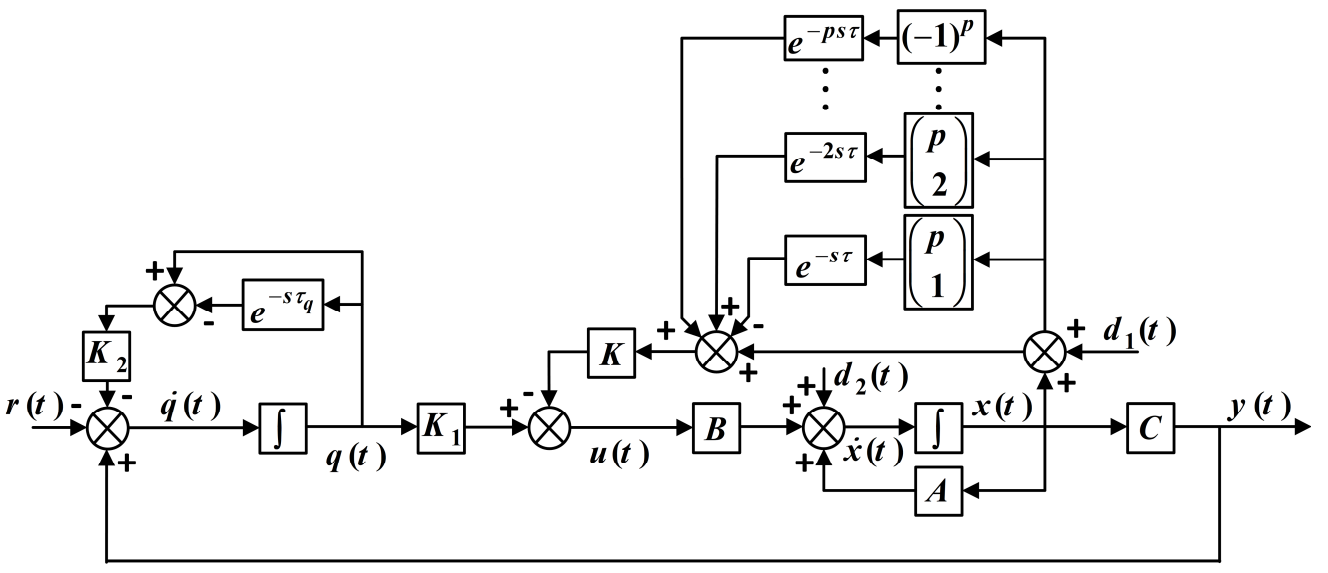

Fig. 2. Block diagram of the proposed delayed feedback tracker with the modified form of the integral state.

Simulated annealing (SA) algorithm introduced by Kirkpatrick [41], is a probabilistic technique to find a global optimum of a function. This algorithm was firstly inspired by statistical mechanics and has been applied to several optimization problems such as traveling salesman problem [42]. Since we are not searching for a precise global minimum of the rightmost eigenvalue of (12), and in most practical applications, decreasing rightmost root below a threshold fulfills our objectives, this algorithm would be a suitable option.

SA algorithm has been widely discussed and used for variety of applications [43], [44], [45], [46], [47]. Accordingly, assumptions and mathematical details of the algorithm will not be discussed herein, and we directly utilize this method for designing delayed feedback controller parameters.

In order to apply SA to our problem, an energy function $E(W)$ must be defined which shows the energy of the system at the state $W$ . In this case, $E(W)$ indicates the rightmost root of the characteristic equation (12) and the state $W$ is the set of the controller parameters or $W \triangleq (K, K_1, K_2, \tau, \tau_q)$. Another important element in SA is the temperature $T$ which should be initialized appropriately for different optimization problems. Finally, the $P$ probability of accepting a state change is determined by the Boltzmann distribution of the energy difference of the two states as follows.

$$P = e^{-\frac{\Delta E}{T}}. \tag{14}$$

By these definitions, we introduce the following SA algorithm for determining the delayed feedback controller parameters $W$ in order to stabilize (9).

***The SA Algorithm for determining the proposed delayed feedback controller parameters***

Step 1. Initializing the controller parameters $W$ by random values.
Step 2. Initializing $T$ with a large value.
Step 3. **Repeat:**

a. **Repeat:**

I. Applying random perturbation to the controller parameters as $W = W + \Delta W$ .

II. Computing $\Delta E = E(W + \Delta W) - E(W)$:

if $\Delta E < 0$ , keep the new set of the controller parameters $W$ .
else accept the new set of the controller parameters with the probability:

$$P = e^{-\frac{\Delta E}{T}}.$$

until the number of the accepted changes is less than a specific threshold level.

b. Set $T = T - \Delta T$

until $T$ is close enough to zero or $E(W)$ reaches a desired value which can be determined by the designer.

In this algorithm, $E(W)$ or the rightmost root of (12) can be calculated effectively using the Matlab package DDE-BIFTOOL [48]. Notice that in each iteration, $\Delta W$ must be chosen according to the physical constraints on the controller parameter values, i.e. delays must be kept positive during the algorithm. Moreover, in [49], it has been shown that if $T$ is decreased such that

$$T(k) \geq \frac{T_0}{\ln(1+k)}, k = 1, 2, ... \tag{15}$$

where $k$ indicates the iteration number and $T_0$ is a sufficiently large initial temperature, the SA algorithm will find the global optimum with probability value one.

## IV. DISTURBANCE REJECTION

In this section, the robustness of the proposed control approach is discussed in the presence of the disturbance signals $d_1(t)$ and $d_2(t)$ shown Fig. 2.

### A. Rejection of disturbance signal $d_1(t)$

In this subsection, the effect of the disturbance signal $d_1(t)$ on the system states is studied. To this end, the disturbance signal $d_1(t)$ is added to the control signal $u(t)$ in (6) and afterwards by considering $r(t)=0$ the closed loop system equations are obtained as below.

$$\begin{cases}\begin{bmatrix}\dot{x}(t)\\ \dot{q}(t)\end{bmatrix}=\begin{bmatrix}A-BK & -BK_1\\ C & -K_2\end{bmatrix}\begin{bmatrix}x(t)\\ q(t)\end{bmatrix}\\ +\begin{bmatrix}-BK & 0\\ 0 & 0\end{bmatrix}\left(\sum_{i=1}^{p}\left(\begin{pmatrix}p\\ i\end{pmatrix}(-1)^i\begin{bmatrix}x(t-i\tau)\\ q(t-i\tau)\end{bmatrix}\right)\right)\\ +\begin{bmatrix}0 & 0\\ 0 & K_2\end{bmatrix}\begin{bmatrix}x(t-\tau_q)\\ q(t-\tau_q)\end{bmatrix}\\ +\begin{bmatrix}-BK & 0\\ 0 & 0\end{bmatrix}\left(\sum_{i=0}^{p}\left(\begin{pmatrix}p\\ i\end{pmatrix}(-1)^i\begin{bmatrix}d_1(t-i\tau)\\ 0\end{bmatrix}\right)\right),\\ y(t)=\begin{bmatrix}C & 0\end{bmatrix}\begin{bmatrix}x(t)\\ q(t)\end{bmatrix}.\end{cases} \tag{16}$$

According to (8), the Laplace transform of $q(t)$, i.e. $Q(s)$ in terms of the Laplace transform of $x(t)$, i.e. $X(s)$ is described as

$$Q(s)=(sI+K_2(1-e^{-s\tau_q}))^{-1}CX(s). \tag{17}$$

In addition, based on the system equations (16) the following equality holds.

$$L\{\dot{x}(t)\}=L\Big\{Ax(t)-BK_1q(t)\\ -BK\left(\sum_{i=0}^{p}\left(\begin{pmatrix}p\\ i\end{pmatrix}(-1)^i x(t-i\tau)\right)\right)\\ -BK\left(\sum_{i=0}^{p}\left(\begin{pmatrix}p\\ i\end{pmatrix}(-1)^i d_1(t-i\tau)\right)\right)\Big\}. \tag{18}$$

Substituting $Q(s)$ from (17) into (18), eventuates to the following equality

$$X(s)=-\Big[sI-A+BK(1-e^{-s\tau})^p\\ +BK_1(sI+K_2(1-e^{-s\tau_q}))^{-1}C\Big]^{-1}BK(1-e^{-s\tau})^p D_1(s). \tag{19}$$

where $D_1(s)$ is the Laplace transform of $d_1(t)$. The equality (19) can be also rewritten in the following form

$$X(s)=-s\Big[s^2I-sA+sBK(1-e^{-s\tau})^p\\ +BK_1(I+K_2(\frac{1-e^{-s\tau_q}}{s}))^{-1}C\Big]^{-1}BK(1-e^{-s\tau})^p D_1(s). \tag{20}$$

Now, suppose $D_1(s)$ affecting all the measured states as follows.

$$D_1(s)=\frac{1}{s^{p+2}}\begin{bmatrix}1\\ \vdots\\ 1\end{bmatrix}_{n\times 1}=\frac{1}{s^{p+2}}V_{n\times 1}, \tag{21}$$

where $V=\begin{bmatrix}1\\ \vdots\\ 1\end{bmatrix}_{n\times 1}$. Substituting $D_1(s)$ in (21) into (20) and applying the final value theorem results in

$$\lim_{s\to 0} sX(s)=\lim_{s\to 0}\Bigg(-s^2\Big[s^2I-sA+sBK(1-e^{-s\tau})^p\\ +BK_1(I+K_2(\frac{1-e^{-s\tau_q}}{s}))^{-1}C\Big]^{-1}BK(1-e^{-s\tau})^p\frac{1}{s^{p+2}}V\Bigg). \tag{22}$$

Noticing that $\lim_{s\to 0}\frac{1-e^{-s\tau}}{s}=\tau\neq 0$, relation (22) is simplified to

$$\lim_{s\to 0}\left(-\left[BK_1(I+K_2\tau_q)^{-1}C\right]^{-1}BK(1-e^{-s\tau})^p\frac{1}{s^p}V\right)=\\ \lim_{s\to 0}\Big(-\left(\det(I+K_2\tau_q)\right)\left[BK_1\mathrm{adj}\left(I+K_2\tau_q\right)C\right]^{-1}\\ BK(1-e^{-s\tau})^p\frac{1}{s^p}V\Big). \tag{23}$$

In (23), adj(.) stands for the adjoint of the matrix. From (23), it can be inferred that the steady state value of $x(t)$ in the presence of $d_1(t)=t^{p+2}V$ equals zero if we have the following conditions

$$\det(I+K_2\tau_q)=0, \tag{24}$$

According to (24), in order to reject the disturbance signal $d_1(t)=t^{p+2}V$, the matrix $K_2\tau_q$ must have some eigenvalues in $-1$. It is worthwhile to state that if (24) is not fulfilled, the closed loop system is able to reject the disturbance signal $d_1(t)=t^{p+1}V$. Benefiting from the control signal $u(t)$ in (6) the proposed control structure is able to reject all the disturbance signals $d_1(t)=t^lV$ for $l\geq 0, l\in\mathbb{Z}$. Moreover, using the integral state $q(t)$ as (8) may improve the disturbance rejection property of the closed loop system by one more order.

*B. Rejection of disturbance signal $d_2(t)$*

In this subsection, it is investigated that for an appropriate choice of the matrix $K_2\tau_q$, the closed loop system rejects the step/ramp type of the disturbance signal $d_2(t)$.

Considering the disturbance signal $d_2(t)$ and $r(t)=0$ the closed loop system equation is obtained as follows.

$$\begin{cases} \begin{bmatrix} \dot{x}(t) \\ \dot{q}(t) \end{bmatrix} = \begin{bmatrix} A-BK & -BK_1 \\ C & -K_2 \end{bmatrix} \begin{bmatrix} x(t) \\ q(t) \end{bmatrix} \\ + \begin{bmatrix} -BK & 0 \\ 0 & 0 \end{bmatrix} \left( \sum_{i=1}^{p} \left( \begin{pmatrix} p \\ i \end{pmatrix} (-1)^i \begin{bmatrix} x(t-i\tau) \\ q(t-i\tau) \end{bmatrix} \right) \right) \\ + \begin{bmatrix} 0 & 0 \\ 0 & K_2 \end{bmatrix} \begin{bmatrix} x(t-\tau_q) \\ q(t-\tau_q) \end{bmatrix} + \begin{bmatrix} I \\ 0 \end{bmatrix} d_2(t), \\ y(t) = \begin{bmatrix} C & 0 \end{bmatrix} \begin{bmatrix} x(t) \\ q(t) \end{bmatrix}. \end{cases} \tag{25}$$

According to (25), the Laplace transform of $x(t)$ is given by

$$X(s) = -\Big[ sI - A + BK(1-e^{-s\tau})^p + BK_1(sI + K_2(1-e^{-s\tau_q}))^{-1}C \Big]^{-1} D_2(s). \tag{26}$$

Similar to (20), $X(s)$ can be rewritten in the following form

$$X(s) = -s\Big[ s^2 I - sA + sBK(1-e^{-s\tau})^p + BK_1(I + K_2(\frac{1-e^{-s\tau_q}}{s}))^{-1}C \Big]^{-1} D_2(s). \tag{27}$$

Suppose the Laplace transform of $d_2(t)$ as $D_2(s) = \frac{1}{s^2} V_{n\times 1}$. Now, applying the finite value theorem results in

$$\lim_{s\to 0} sX(s) = \left( -\left(\det(I + K_2\tau_q)\right)\left[ BK_1 \mathrm{adj}\left(I + K_2\tau_q\right)C \right]^{-1} V \right). \tag{28}$$

From (28) it is induced that the steady state value of $x(t)$ equals zero i.e. the step/ramp type disturbance $d_2(t)$ is rejected if the equality (24) is fulfilled.

From the detailed discussions in this section, it is concluded that the proposed control method in this paper is robust against the disturbance signal $d_1(t)$ of any order showing in Fig. 2. Moreover, it is robust against the step/ramp type disturbance signal $d_2(t)$ in Fig. 2. Thus, the proposed control system improves the disturbance rejection characteristic of the conventional state feedback by applying simple and implementionally convenient delayed feedback control.

## V. Tracking Step and Ramp Type Reference Inputs

In this section, it is shown that satisfying a specific condition the proposed control system in this paper precisely tracks the ramp type reference inputs as well as the step type inputs.

According to (9) and considering $Q(s)$ as the Laplace transform of $q(t)$ the following equality is resulted.

$$X(s) = -\left[ sI - A + BK(1-e^{-s\tau})^p \right]^{-1} BK_1 Q(s). \tag{29}$$

From (8) and (29) it is concluded that

$$Q(s) = -\Big[ sI + C\left[ sI - A + BK(1-e^{-s\tau})^p \right]^{-1} BK_1 + K_2(1-e^{-s\tau_q}) \Big]^{-1} R(s). \tag{30}$$

Moreover from (8) and according to Fig. 2, the Laplace transform of the tracking error $e(t) = y(t) - r(t)$ is given by

$$E(s) = \left[ sI + K_2(1-e^{-s\tau_q}) \right] Q(s). \tag{31}$$

Substituting $Q(s)$ from (30) into (31) results in

$$E(s) = \Big[ sI + K_2(1-e^{-s\tau_q}) \Big] \Big[ sI + K_2(1-e^{-s\tau_q}) + C\left[ sI - A + BK(1-e^{-s\tau})^p \right]^{-1} BK_1 \Big]^{-1} R(s). \tag{32}$$

Suppose the Laplace transform of the reference input $r(t)$ as $R(s) = \frac{1}{s^k} M_{m\times 1}$ where $M_{m\times 1} = \begin{bmatrix} 1 \\ \vdots \\ 1 \end{bmatrix}_{m\times 1}$. Applying the final value theorem yields to the following equality

$$\begin{aligned} \lim_{s\to 0} sE(s) = \lim_{s\to 0} s^2 \Bigg( \left( I + K_2(\frac{1-e^{-s\tau_q}}{s}) \right) \\ \left( sI + C\left( sI - A + BK(1-e^{-s\tau})^p \right)^{-1} BK_1 \right. \\ \left. + K_2(1-e^{-s\tau_q}) \right)^{-1} \frac{1}{s^k} M_{m\times 1} \Bigg) = \\ \lim_{s\to 0} \left( \left(I + K_2\tau_q\right)\left(C(-A)^{-1}BK_1\right)^{-1} \frac{1}{s^{k-2}} M_{m\times 1} \right). \end{aligned} \tag{33}$$

From (33), it is inferred that for $k=2$ i.e. the ramp reference input, the steady state value of the tracking error equals zero if the equality $I+K_2\tau_q=0$ is satisfied. Also, from (33) it is obvious that tracking the reference input is definitely achieved for $k=1$ i.e. for the step type reference input.

## VI. CASE STUDY

In this section, the efficiency of the proposed control approach in reference tracking and disturbance rejection is shown on a case study. Consider the state space matrices of the controllable system (1) are given as follows.

$$A=\begin{bmatrix}3 & -3.75\\ 1 & -1\end{bmatrix},\ B=\begin{bmatrix}1\\ -1.5\end{bmatrix},\ C=[-2.5\quad 2],$$

where the matrix $A$ has two eigenvalues 0.5 and 1.5 in the right half plane.

Considering $u(t)$ in (6) for $p=1$ and the integral state as (8) for $\tau_q=0$, the SA algorithm is used to find the appropriate values of the controller parameters $K=\begin{bmatrix}k_1 & k_2\end{bmatrix}$, $K_1=k_{11}$ and $\tau$ in (6). Fig. 3a indicates the delayed feedback controller parameters with respect to the iteration number of the SA algorithm. Furthermore, Fig. 3b shows the real part of the rightmost root of the characteristic equation (12) with respect to the iteration number of the SA algorithm.

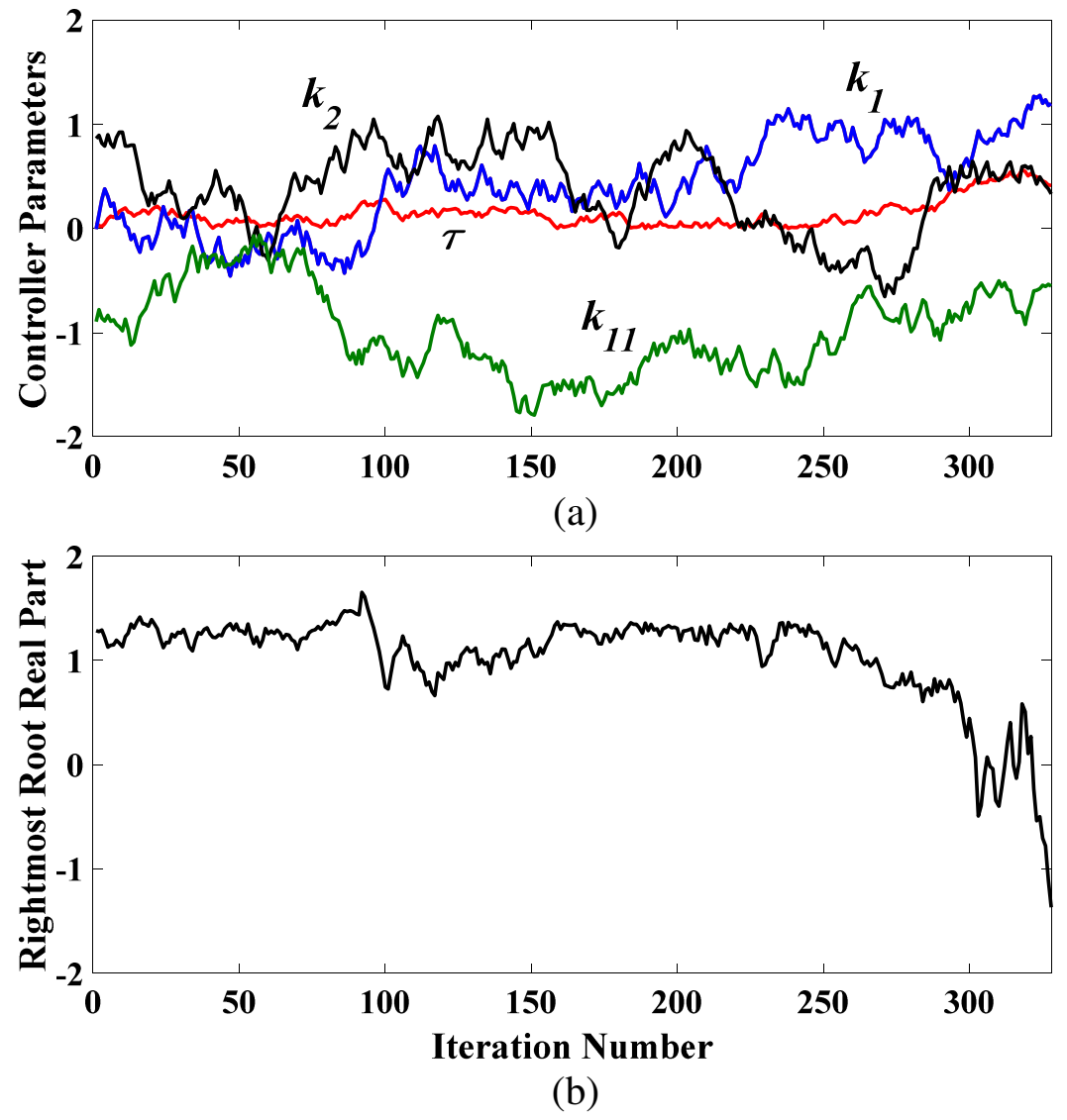

Fig. 3. Designing the delayed feedback controller parameters with SA algorithm.
(a) Delayed feedback controller parameters $K=\begin{bmatrix}k_1 & k_2\end{bmatrix}$, $K_1=k_{11}$ and $\tau$ with respect to the iteration number of the SA algorithm for $p=1$ and $\tau_q=0$.
(b) The real part of the rightmost root of the characteristic equation (12) with respect to the iteration number of the SA algorithm for $p=1$ and $\tau_q=0$.

As it is seen in these figures, the SA algorithm stops when the real part of the rightmost root becomes less than a predefined threshold which is considered to be equal -1 in this example. According to the obtained results the values of these parameters are chosen as $K=\begin{bmatrix}1.2 & 0.3319\end{bmatrix}$, $K_1=-0.5523$, and $\tau=0.41$. The disturbance signal $d_1(t)$ on the system sates is shown in Fig. 4a. As this figure depicts the disturbance signal $d_1(t)$ on the system sate $x_1(t)$ is applied as a combination of step and ramp signals. Outputs of the closed loop system by applying the conventional state feedback method and the proposed delayed feedback method in the presence of the disturbance signal $d_1(t)$ are shown in Fig. 4b.

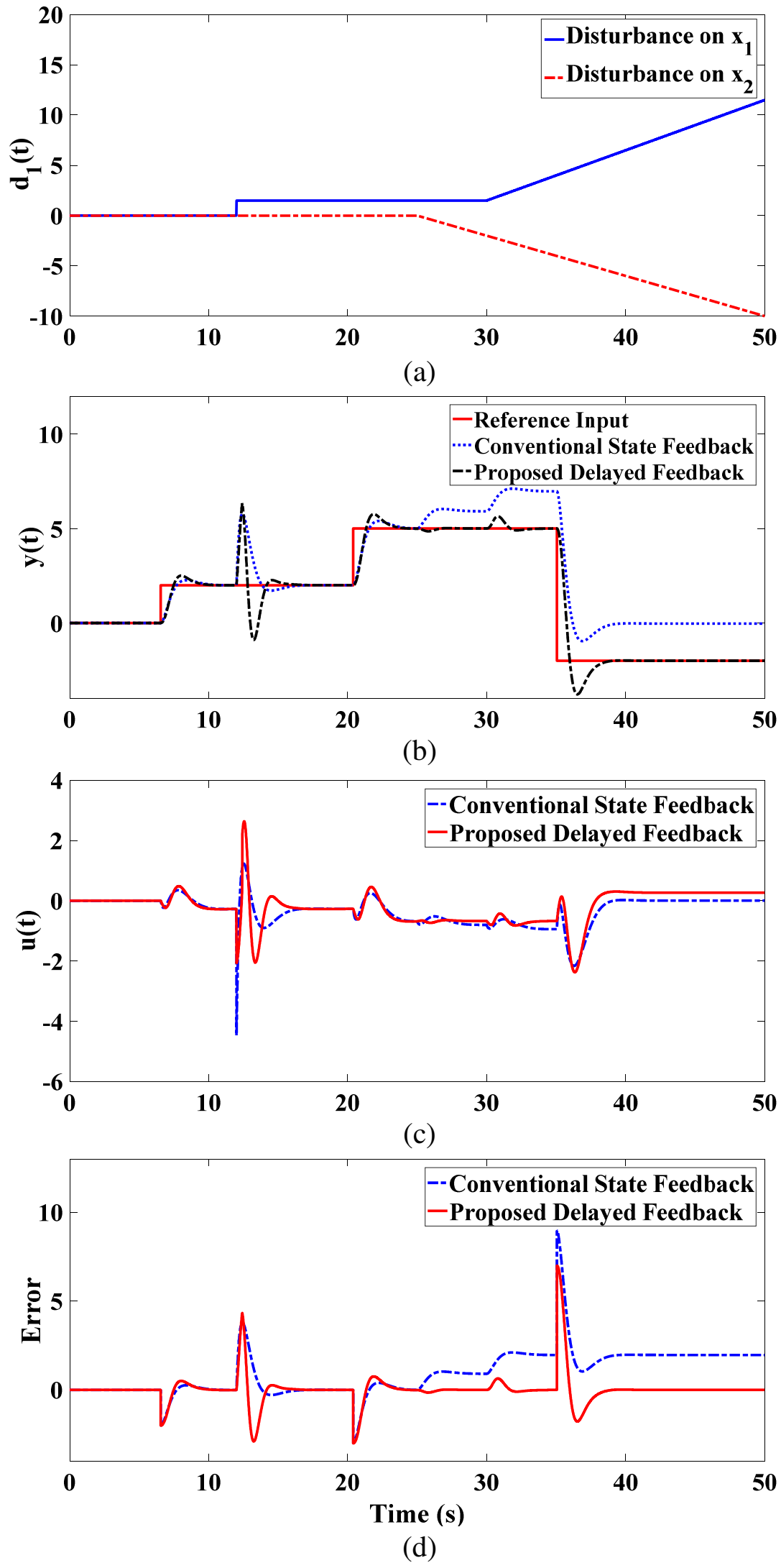

Fig. 4. The corresponding signals of the proposed delayed feedback with $p=1$ and $\tau_q=0$ and the conventional state feedback with step reference input.
(a) Disturbance signal $d_1(t)$ imposed on the system states $x_1(t)$ and $x_2(t)$.
(b) Comparison of the closed loop system outputs by applying the conventional state feedback and the proposed delayed feedback method in the presence of the disturbance signal $d_1(t)$ in Fig. 4a.
(c) Control signal $u(t)$ corresponding to Fig. 4b.
(d) Tracking error signals corresponding to Fig. 4b.

Moreover, according to the Fig. 4c, the control signal $u(t)$ of the proposed delayed feedback remains bounded during the tracking process. Also, Fig. 4d indicates the error signals corresponding to Fig. 4b. As it is seen in these figures, the conventional state feedback method cannot reject the disturbance signal $d_1(t)$, while the proposed delayed feedback method omits the effect of the disturbance signal $d_1(t)$ in the closed loop system output completely. It is worth mentioning that to make a fair comparison, three roots of the characteristic equation by applying the conventional state feedback method has been located in the same place of the three dominant roots of the characteristic equation by applying the proposed delayed feedback method. These places are $-1.36 \pm 0.9646i, -2.6729$.

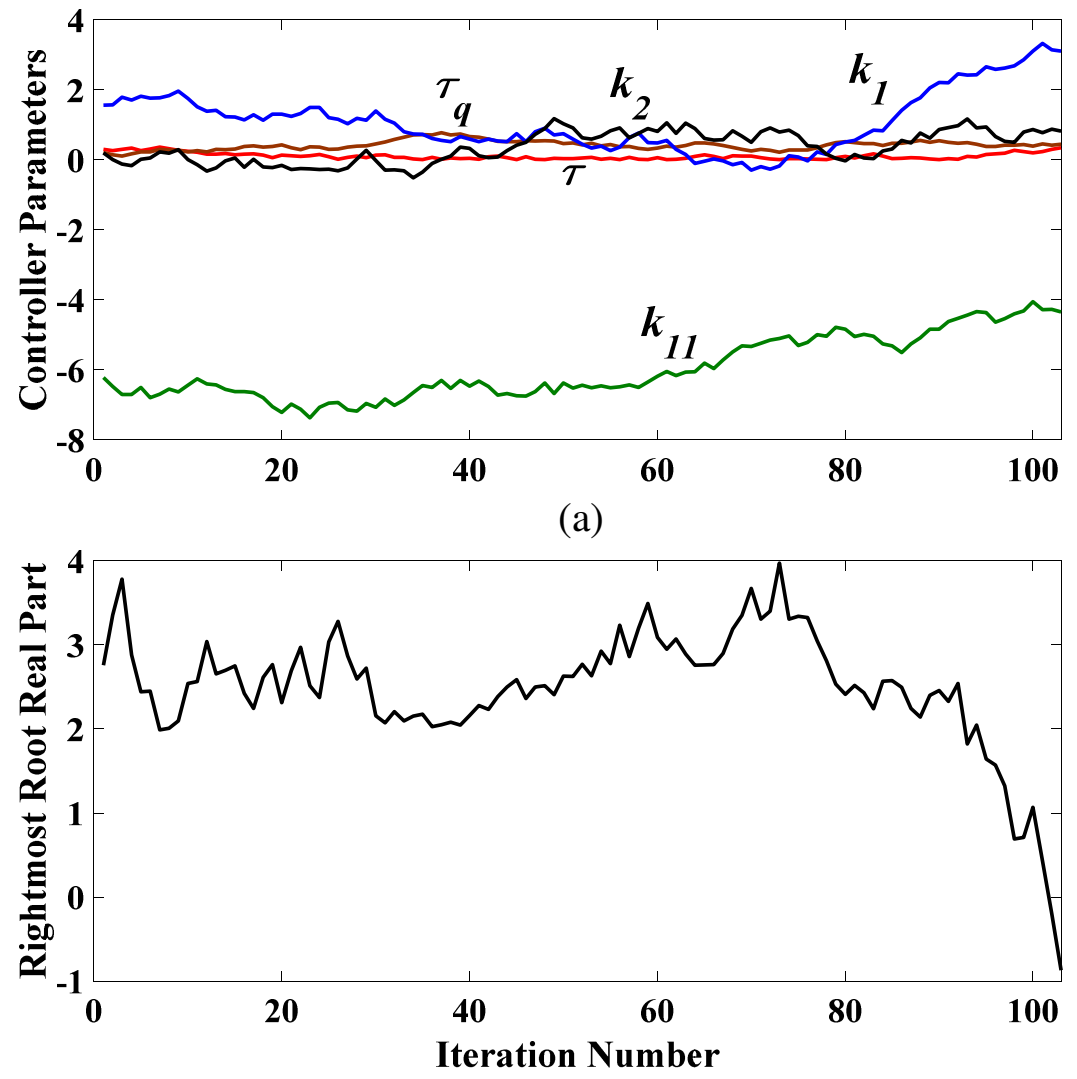

Fig. 5. Designing the delayed feedback controller parameters with SA algorithm.
(a) Delayed feedback controller parameters $K = [k_1 \quad k_2]$, $K_1 = k_{11}$, $\tau$ and $\tau_q$ with respect to the iteration number of the SA algorithm for $p = 1$ and $\tau_q \neq 0$.
(b) The rightmost root of the characteristic equation (12) with respect to the iteration number of the SA algorithm for $p = 1$ and $\tau_q \neq 0$.

In the following, the delayed feedback signal $u(t)$ in (6) is considered with $p = 1$ and $\tau_q \neq 0$. Moreover, $K_2 = -1/\tau_q$ is considered. Similar to the previous design, the SA algorithm is used to obtain the other control parameters $K = [k_1 \quad k_2]$, $K_1 = k_{11}$ and $\tau$ in (6) and $\tau_q$ in (8). According to the obtained result in Fig. 5a the value of these parameters are chosen as $K = [3.097 \quad 0.8184]$, $K_1 = -4.346$, $\tau = 0.34$ and $\tau_q = 0.44$. Fig. 5b also shows the real part of the rightmost root of the characteristic equation (12) with respect to the iteration number of the SA algorithm. Fig. 6a and Fig. 6b depicts the disturbance signals $d_1(t)$ and $d_2(t)$ on the system states. As it is seen, the disturbance signal $d_1(t)$ is a parabolic signal and the disturbance signal $d_2(t)$ is a combination of step and ramp signals. Output of the closed loop system by applying the proposed delayed feedback method in the presence of the simultaneous disturbance signals $d_1(t)$ and $d_2(t)$ is represented in Fig. 7a. In addition, Fig. 7b implies that under these circumstances, the control signal $u(t)$ of the proposed delayed feedback still remains bounded.

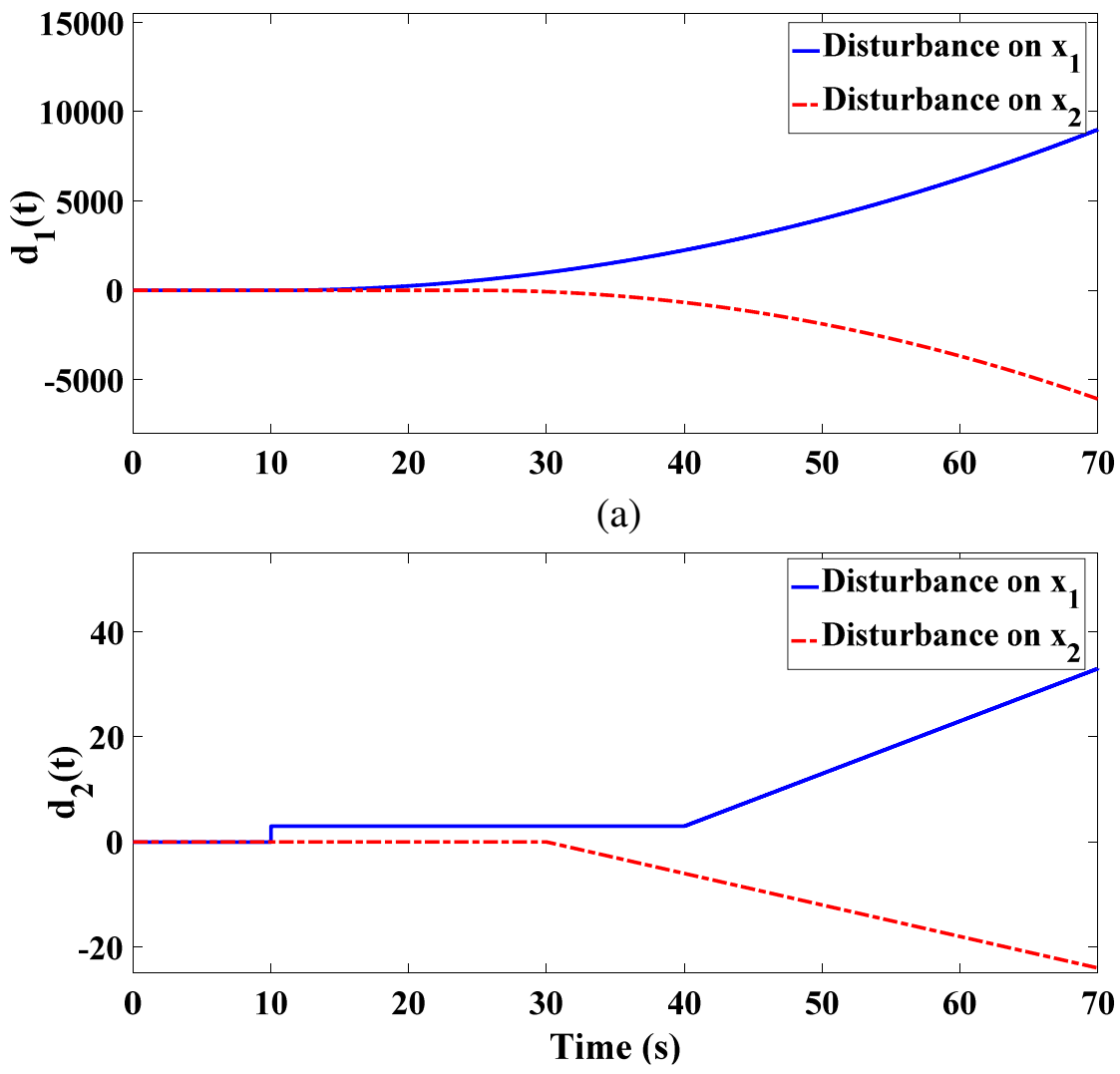

Fig. 6. Disturbance signals imposed on the closed loop system in Fig. 2.
(a) $d_1(t)$ parabolic disturbance signal imposed on $x_1(t)$ and $x_2(t)$ states.
(b) $d_2(t)$ disturbance signal imposed on $x_1(t)$ and $x_2(t)$ states.

Also, the tracking error of the proposed control system in this condition is depicted in Fig. 7c. As it is seen, the proposed system is able to track the reference input signal and reject both of the disturbances in the steady state without any error. Moreover, as it is obvious from Fig. 7d, the conventional state feedback method cannot handle such a combination of disturbance signals. Therefore, other corresponding signals have not been compared in this case.

Finally, Fig. 8a shows the reference input tracking of the designed control system by applying a combination of step and ramp reference input signals and also imposing simultaneous disturbance signals $d_1(t)$ and $d_2(t)$. Likewise, it is seen from Fig. 8b that the control signal $u(t)$ is bounded. Also, the tracking error under these circumstances is presented in Fig. 8c. As what was proved in Section 4, this sophisticated combination of reference input and disturbance signals can be also effectively handled by the design control system based on the delayed feedback method. Whereas, Fig. 8 shows that under these circumstances, the conventional state feedback is completely unable to track the reference input signal.

In summary, the comparison between the proposed delayed feedback and the conventional state feedback tracking methods is given in Table I.

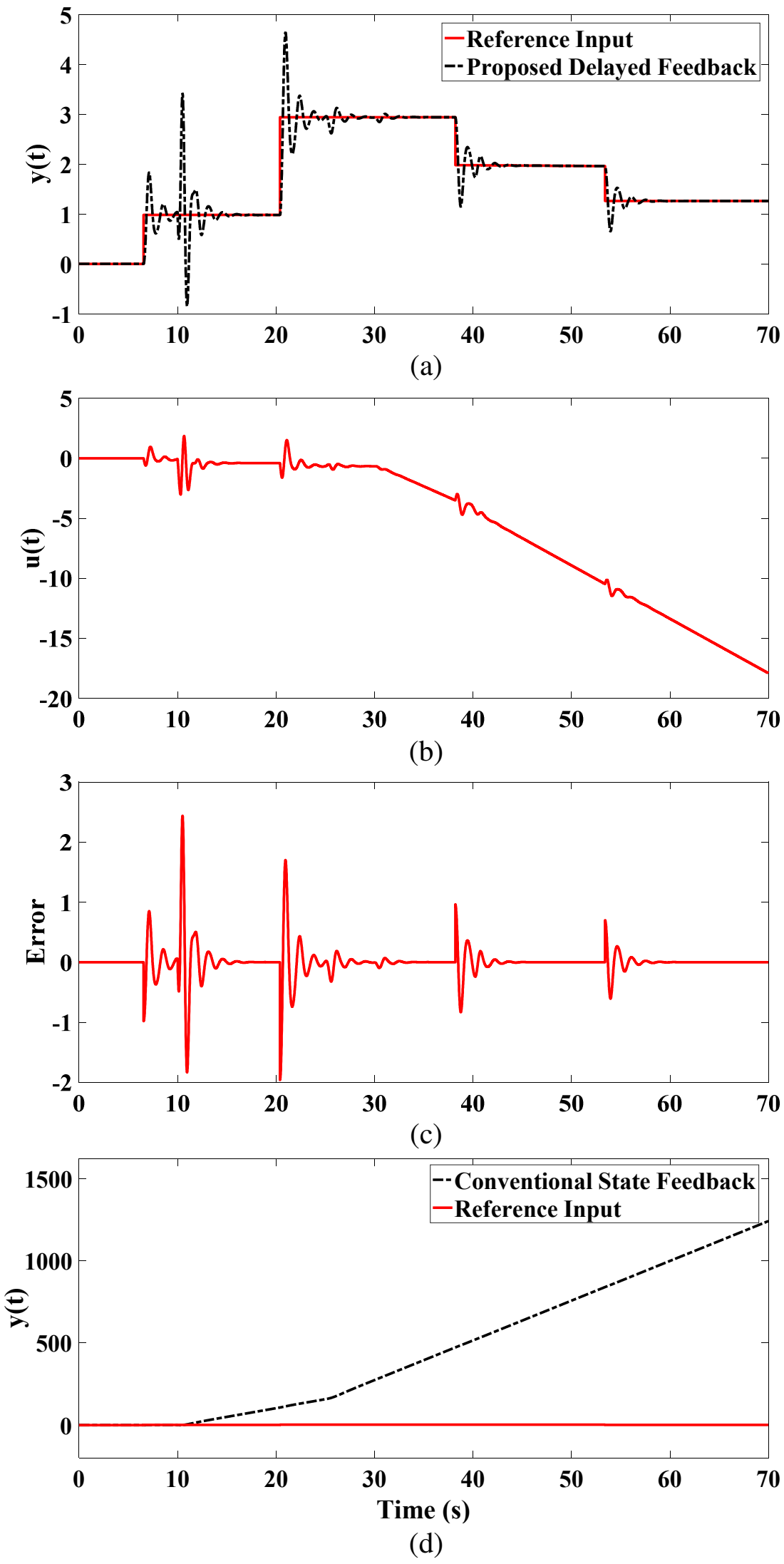

Fig. 7. The corresponding signals of the proposed delayed feedback with $p = 1$ and $\tau_q \neq 0$ and output of the conventional state feedback with step reference input.
(a) Closed loop system output with the proposed delayed feedback in the presence of the simultaneous disturbance signals $d_1(t)$ in Fig. 6a and $d_2(t)$ in Fig. 6b.
(b) Control signal $u(t)$ corresponding to Fig. 7a.
(c) Tracking error signal corresponding to Fig. 7a.
(d) Closed loop system output with the conventional state feedback in the presence of the simultaneous disturbance signals $d_1(t)$ in Fig. 6a and $d_2(t)$ in Fig. 6b.

## VII. Conclusion

In this study, the conventional integral action tracking method has been improved by the delayed feedback control scheme. The proposed method stabilizes the conventional integral action tracking system using the delayed feedback. In this way, it tries to integrate the delayed feedback features with the conventional tracking system structure. Due to the delayed feedback characteristics, the proposed method is able to reject higher order disturbances. Besides, based on the non-invasive feature of the delayed feedback, a new term is added to the integral state which increases the closed loop system disturbance rejection property by one order. Moreover, by satisfying a simple condition, this new term will cause the closed loop system to track step/ramp reference input signals. The proposed system possesses a convenient implementation feature which makes it a suitable choice for most practical applications.

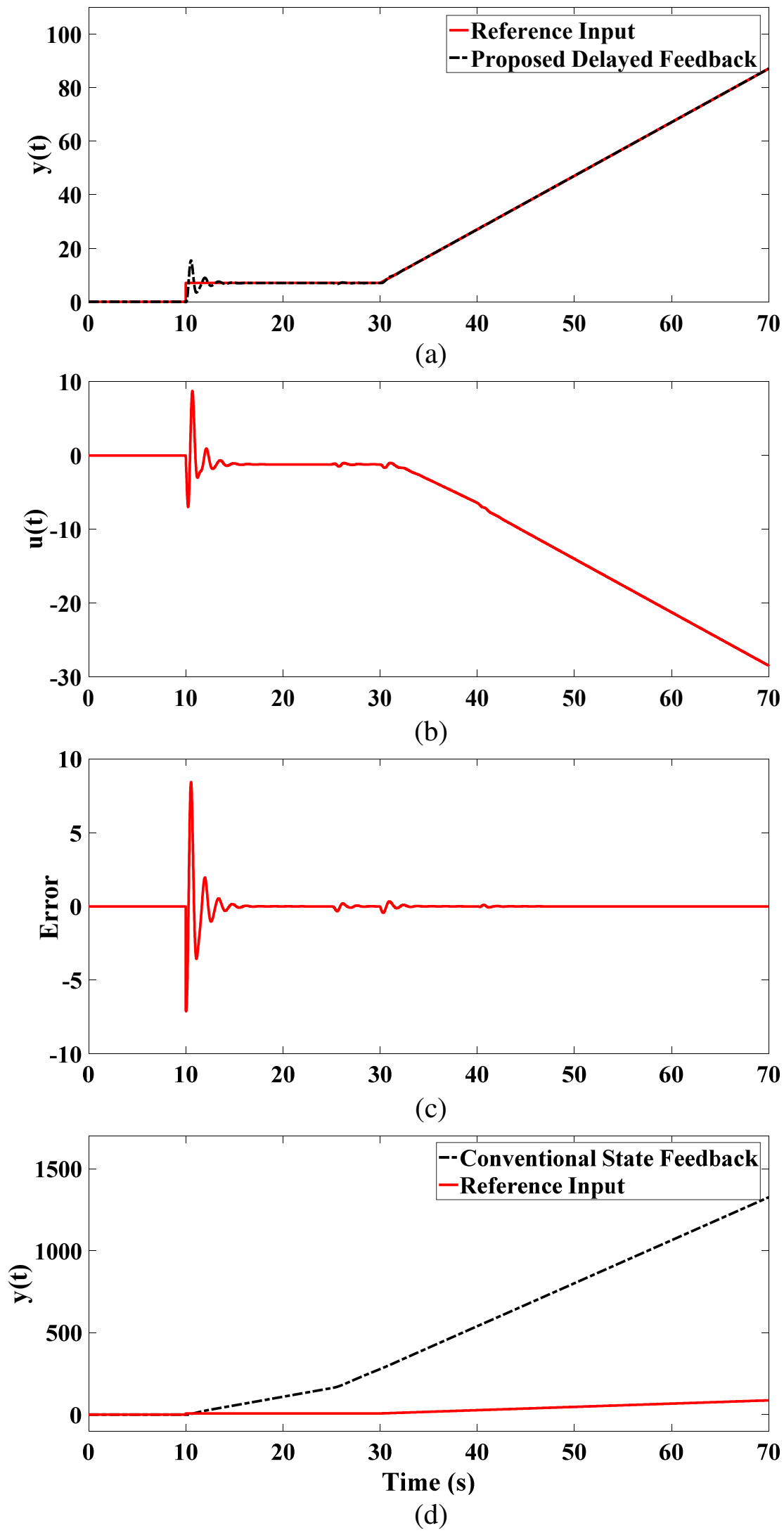

Fig. 8. The corresponding signals of the proposed delayed feedback with $p = 1$ and $\tau_q \neq 0$ and output of the conventional state feedback with a combination of ramp and step reference input.
(a) Closed loop system output with the proposed delayed feedback by applying step and ramp reference inputs and in the presence of the simultaneous disturbance signals $d_1(t)$ in Fig. 6a and $d_2(t)$ in Fig. 6b .
(b) Control signal $u(t)$ corresponding to Fig. 8a.
(c) Tracking error signal corresponding to Fig. 8a.
(d) Closed loop system output with the conventional state feedback by applying step and ramp reference inputs and in the presence of the simultaneous disturbance signals $d_1(t)$ in Fig. 6a and $d_2(t)$ in Fig. 6b.

TABLE I
COMPARISON BETWEEN THE PROPOSED DELAYED FEEDBACK AND THE CONVENTIONAL STATE FEEDBACK TRACKING METHODS.

| Order | One | Two | Higher |
|---|---|---|---|
| $d_1(t)$ rejection by the proposed delayed feedback method | Yes | Yes | Yes |
| $d_1(t)$ rejection by the conventional state feedback method | Yes | No | No |
| $d_2(t)$ rejection by the proposed delayed feedback method | Yes | Yes | No |
| $d_2(t)$ rejection by the conventional state feedback method | Yes | No | No |
| Reference input tracking by the proposed delayed feedback method | Yes | Yes | No |
| Reference input tracking of the conventional state feedback | Yes | No | No |